\documentclass[twocolumn,superscriptaddress,longbibliography,amsmath,amssymb,aps,pra]{revtex4-1}
\pdfoutput=1
\usepackage{graphicx}
\usepackage{dcolumn}
\usepackage{bm}

\begin{document}
\newcommand{\comment}[1]{}

\title{Multiphoton-Excited Fluorescence of Silicon-Vacancy Color Centers in Diamond }

\author{J.~M.~Higbie}\email{jhigbie@google.com}
\author{ J.~D.~Perreault }
 \affiliation{%
 Verily Life Sciences, Mountain View, CA
 }%

\author{ V.~M.~Acosta
}
\affiliation{%
University of New Mexico, Albuquerque, NM
}%

\author{ C.~Belthangady}
 \affiliation{%
 Verily Life Sciences, Mountain View, CA
 }%

\author{ P.~Lebel}
\email[Current address: Berkeley Lights, Emeryville, CA]{
 }%

 \author{M.~H.~Kim}
\author{K.~Nguyen}
\author{ V.~Demas}
\author{ V.~Bajaj}
\author{ C.~Santori}
 \affiliation{%
 Verily Life Sciences, Mountain View, CA
 }%

%
%



\date{\today}

\begin{abstract}
  Silicon-vacancy  color  centers  in nanodiamonds  are  promising  as
  fluorescent  labels  for  biological applications,  with  a  narrow,
  non-bleaching  emission line  at 738\,nm.  Two-photon excitation  of
  this fluorescence offers the possibility of low-background detection
  at  significant tissue  depth  with  high three-dimensional  spatial
  resolution.   We have  measured  the  two-photon fluorescence  cross
  section  of  a  negatively-charged   silicon  vacancy  (SiV$^-$)  in
  ion-implanted          bulk          diamond          to          be
  $0.74(19)  \times  10^{-50}{\rm  cm^4\;s/photon}$ at  an  excitation
  wavelength  of  1040\,nm.  In  comparison  to  the diamond  nitrogen
  vacancy  (NV)   center,  the  expected  detection   threshold  of  a
  two-photon excited  SiV center  is more than  an order  of magnitude
  lower, largely due to its much narrower linewidth.
We also present   measurements of two- and three-photon
excitation  spectra, finding an increase in the two-photon cross section with decreasing wavelength,
and  discuss the physical interpretation of the spectra
in the context of  existing models of the SiV energy-level structure.

\end{abstract}


\maketitle

\section{Introduction}
Color centers in diamond 
have  been  the  focus  of  intense interest  in  recent  years.   The
nitrogen-vacancy (NV) color center in  diamond has driven much of this
interest,   thanks  to   numerous  promising   applications  including
nano-scale magnetometry
\cite{balasubramanian2008nanoscale,maze2008nanoscale,
taylor2008high,acosta2010broadband,
maletinsky2012robust},                
NMR spectroscopy
\cite{mamin2013nanoscale,staudacher2013nuclear}, 
quantum information
\cite{dutt2007quantum,
neumann2008multipartite,santori2010nanophotonics,
fuchs2011quantum,faraon2012coupling,childress2013diamond},
 and use  as  fluorescent bio-labels
 \cite{vaijayanthimala2012long,igarashi2012real}.
More recently, the  silicon-vacancy (SiV)
 color center  in diamond, which has been known for over two decades
 \cite{sternschulte19941,clark1995silicon,goss1996twelve},
has generated increasing   excitement
 \cite{rogers2014electronic,neu2013low,neu2012photophysics,hepp2014electronic}
because of  properties that  are in some  respects even  more favorable
than  those  of   the  nitrogen-vacancy  defect,  such   as  a  narrow
zero-phonon   line  (ZPL) \cite{sternschulte19941}   and  weak   phonon
side-bands  at  room  temperature \cite{collins1994annealing}.   This
concentrated emission in a narrow ZPL allows
 detection  of silicon-vacancy-containing diamonds at
higher  signal-to-noise  ratios,  and   raises  the  possibility  that
silicon-vacancy-doped   nanodiamonds     bound to specific biomolecular targets 
could be detectable  in the presence
of  high  autofluorescence  background,  e.g. in  deep  and/or  highly
scattering  biological  tissue.

\begin{figure}[h!]
\includegraphics[width=8.6cm]{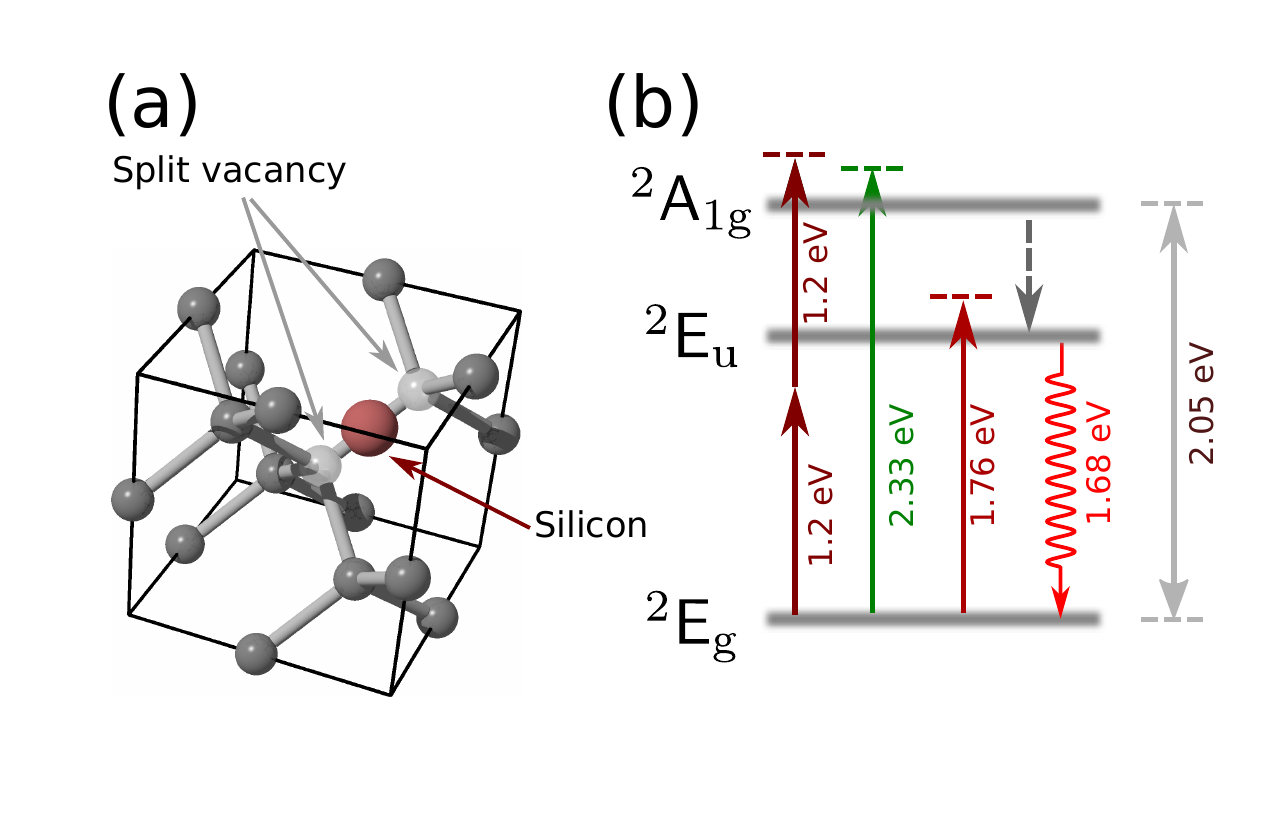}
\caption{\label{fig:latticeLevelsCombined} Diamond unit cell 
containing a  silicon-vacancy
lattice defect (a), 
consisting of a silicon nucleus midway between two
vacant nearest-neighbor lattice sites;
the simplified  level diagram (b) of the negatively charged silicon-vacancy color
center in diamond, showing states and optical transitions relevant to this work,
with 
excitation via two-photon absorption at 1040\,nm (1.19\,eV) or one-photon absorption
at either 705\,nm (1.76\,eV) or 532\,nm (2.33\,eV).}
\end{figure}


Specifically targeted nanoparticle probes that can be detected through
millimeters of intervening tissue promise to be an enabling technology
for  minimally invasive in-vivo  molecular  imaging,   with  potential
applications  in  biomarker
discovery,  studies of immune cell trafficking and circulating tumor cells,
elucidation of molecular pathways in  pre-clinical models,  
drug development, 
and possibly  ultimately  in
clinical diagnosis.     
Towards these ends, various  combinations  of
nanoparticle  type  and  detection modality  have  been  investigated,
including superparamagnetic nanoparticles detected  via MRI \cite{lee2007artificially},  
fluorescence imaging of dyes \cite{ntziachristos2003fluorescence},
quantum dots \cite{gao2005vivo,michalet2005quantum}, and
  nanodiamonds \cite{chung2006spectroscopic,mohan2010vivo}, and  surface-enhanced  Raman
(SERS) particles \cite{qian2008vivo,zavaleta2009multiplexed}.


Two-photon excited fluorescence imaging  is particularly appealing for
deep-tissue imaging  because of  its high spatial  resolution, natural
longitudinal  sectioning,  low  background,  and  because  the  longer
excitation wavelengths  typically used allow enhanced  penetration and
lower   phototoxicity   in  tissue   \cite{helmchen2005deep}.    These
advantages have  led to significant application  of two-photon imaging
in  neuroscience  \cite{svoboda2006principles},  and  will  likely  be
important  in other  areas involving  high spatial-resolution  imaging
through scattering  tissue.  While  two-photon labels such  as organic
dyes   and  fluorescent   proteins   achieve   very  high   brightness
\cite{xu1996measurement,drobizhev2011two},  SiVs are  likely to  offer
distinct advantages thanks to their narrow emission bandwidth and lack
of bleaching.  A  further unique feature of  the silicon-vacancy color
center  is the  circumstance that  the  excitation can  be within  the
second near-IR transmission window of tissue (around 1040\,nm) and the
emission (around  738\,nm) within the first  transmission window (see,
e.g.,      Ref.~\cite{       smith2009bioimaging}).       Song      et
al.\cite{song2014plasmon}   have  previously   observed  a   quadratic
dependence of 830\,nm-excited silicon vacancies coupled to the plasmon
resonances  of   gold  nano-ellipsoids,  attributable   to  two-photon
excitation.    A    quantitative   measure   of   the    strength   of
two-photon-excited  fluorescence,  however,  has not  previously  been
obtained for the SiV defect.  Here  we present the first report of the
two-photon  fluorescence  cross   section  of  the  negatively-charged
silicon  vacancy  (SiV$^-$) color  center  in  diamond. Based  on  the
results, we  evaluate the  prospects of SiV nanodiamond as a  contrast agent  for 
biological labeling  applications, in  particular  for \emph{in-vivo}  and deep-tissue
imaging, where strong background fluorescence and low photon collection
efficiency must  be overcome.

\section{Background}

The structure  of the SiV$^-$ color  center has been elucidated  in recent
years         through          a         number          of
contributions \cite{goss1996twelve,batalov2009low,neu2013low,hepp2014electronic,rogers2014electronic}. 
The color center   consists of  a single silicon atom
located midway  between two adjacent vacant carbon  lattice sites (vacancies),
as shown in Fig.~\ref{fig:latticeLevelsCombined}; this configuration
possesses trigonal  $D_{3d}$ symmetry.  A  level diagram tentatively suggested by
   Rogers  et   al. \cite{rogers2014electronic}  is   given  in
Fig.~\ref{fig:latticeLevelsCombined},  showing  ground  and  excited
states relevant to the present work. Excitation of the defect from the
$^2E_\text{g}$ ground state to the $^2E_\text{u}$ excited state can be
achieved either  directly or  else indirectly via higher-lying levels such as the
 $^2A_\text{1g}$
state, combined with   subsequent (presumed non-radiative) relaxation.  The defect
then returns to  the ground state by emission of  a 1.68\,eV (738\,nm)
photon.  An unusually large portion of the emitted radiation is in the
narrow  zero-phonon line,  improving  the  signal-to-noise ratio  with
which the color center can be detected.

The two-photon fluorescence cross section $\sigma_{2p}$ for a single point fluorophore is defined
by the relation 
\begin{equation}
\label{eq:fundamentalSigmaDef}
 \langle\Gamma \rangle = \sigma_{2p} \langle I^2\rangle
,\end{equation}
where $\Gamma$ is the 
fluorescence photon emission rate, $I$ is the excitation intensity 
at the fluorophore, conventionally measured in units of photon number per area per time, 
and angle brackets indicate time averaging over 
an interval significantly greater than the excited-state decay time.
In practice, one measures not the total emission rate, but the quantity $\Gamma_{\rm det}\equiv
\eta_{\rm det}\langle\Gamma\rangle$, where the detection 
efficiency $\eta_{\rm det}$ includes the fraction of light
collected by the microscope objective and the transmission  of all optical components in the detection path.

Because  a pulsed laser must typically be used to obtain appreciable two-photon-excited
fluorescence, we must further relate the mean square intensity appearing in Eq.~\ref{eq:fundamentalSigmaDef}
to the more readily measurable mean intensity. Specifically, we 
 consider excitation 
via a periodic pulse train with a  temporal profile that is
  well approximated by a sum of gaussian-envelope pulses,
\begin{equation}
I_0(t)= I_\text{peak}\sum_{n=-\infty}^{\infty} \exp(-(t-nT_\text{rep})^2/\tau^2)
,\end{equation}
where $I_0$ is the intensity at the  the laser focus,
$I_\text{peak}$ is the temporal peak intensity, $T_\text{rep}$ is the repetition period,  and 
$\tau$ defines the pulse width.
The time average of the intensity and the square intensity over a time
long compared to the repetition period,
neglecting overlap of distinct pulses, are
readily calculated by integration, and obey the relation
\begin{align}
\label{eq:timeAverageResult}
\left \langle
I_0^2
\right \rangle
&=
\left \langle
I_0
\right \rangle^2
\frac{T_\text{rep}}{\tau\sqrt{2\pi}}
\end{align}
Moreover, the intensity $I_0$ at the laser focus is related to the total laser power by
\begin{align}
\label{eq:powerIntensityRelation}
P &= \int \text{d}x\,\text{d}y\, I(x,y)\equiv I_0 A_\text{ex.}.
\end{align}
Here $A_\text{ex.}$ is the area defined by the  excitation point-spread function
$F_\text{ex.}(x,y)$, i.e.,
\begin{equation}A_\text{ex.}\equiv
\int \text{d}x\,\text{d}y\,
F_\text{ex.}(x,y)
,\end{equation}
with the unit-maximum point-spread function $F_\text{ex.}(x,y)$ in turn defined by the relation,
$I(x,y) = I_0 F_\text{ex.}(x,y)$.

Thus for such a pulse train, the average detection rate of photons emitted by a single color center at the focus of the laser beam is given by
\begin{align}
\Gamma_\text{det,2p}&=
\eta_\text{det}    \sigma_{2p} \langle I_0^2 \rangle\\
&= \eta_\text{det}    \sigma_{2p} \langle I_0\rangle^2 \frac{T_\text{rep}}{\tau \sqrt{2\pi}},
\end{align}
where the time-average intensity at the focus
is given in terms of the time-average excitation power $\langle P \rangle$
by $\langle I_0 \rangle  = \langle P \rangle/A_\text{ex.}$.
For an isolated color center, then, the two-photon cross section is related to measurable 
quantities by the equation
\begin{equation}
\label{eq:sigma2pMeasurablesSingles}
\Gamma_\text{det,2p}=
\eta_\text{det}    \sigma_{2p} \left (\frac{ \langle P \rangle}{A_\text{ex.}}\right)^2  
\frac{T_\text{rep}}{\tau\sqrt{2\pi}} S_P,
\end{equation}
where $S_P$ is an empirically determined saturation factor (described more fully below) accounting for any 
deviation of the intensity dependence from the strictly quadratic dependence of 
Eq.~\ref{eq:fundamentalSigmaDef}.

\begin{figure}
\includegraphics[width=8.6cm]{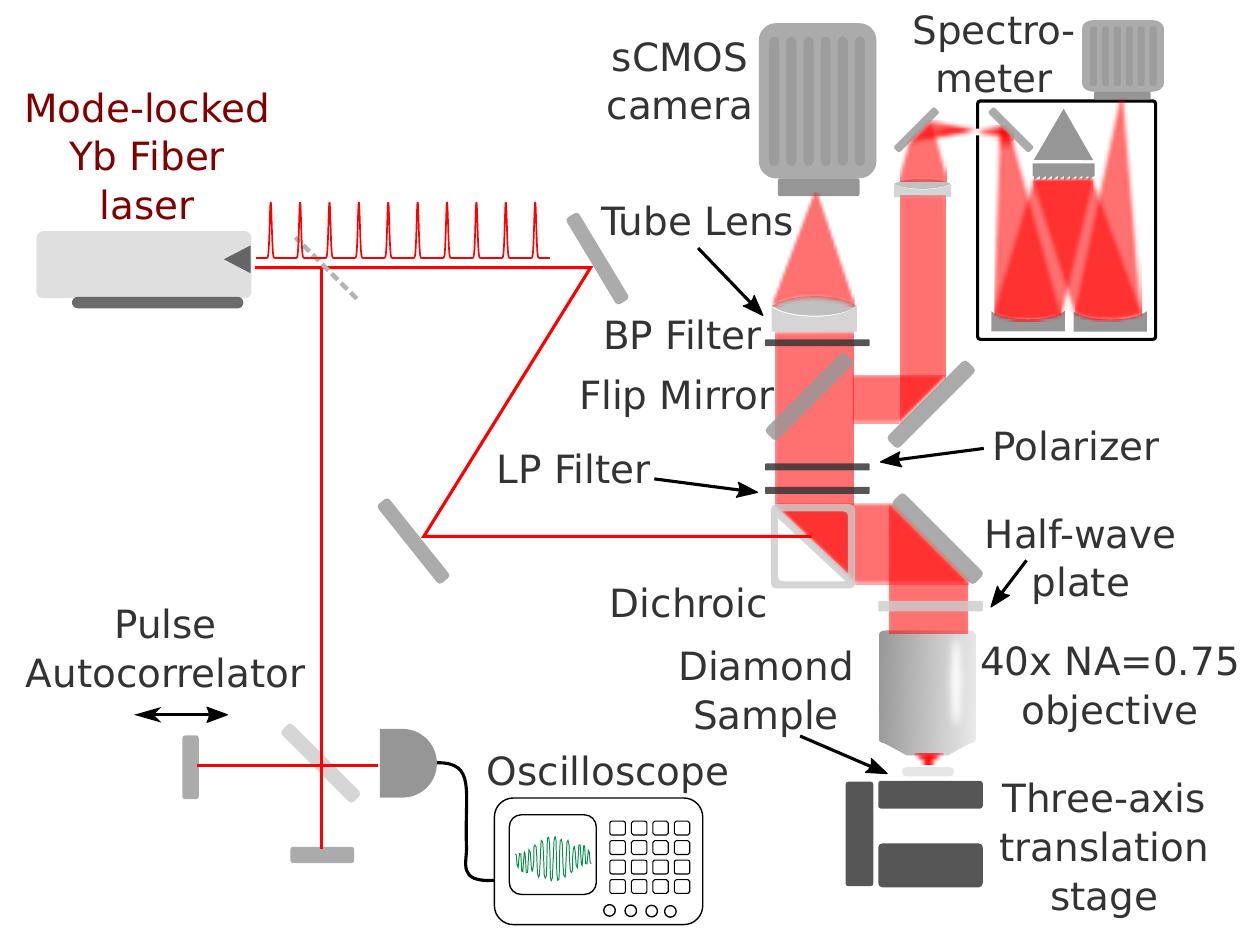}
\caption{\label{fig:apparatusSeparate}
Diagram of 
the two-photon epifluorescence apparatus showing the diamond sample 
 illuminated by the pulsed two-photon excitation laser beam at 1040\,nm or 
by one of two one-photon excitation laser beams, 
with detection by an sCMOS camera or an imaging spectrometer.
An autocorrelator allows measurement of the two-photon laser pulse width. }
\end{figure}

\begin{figure}
\includegraphics[width=8.6cm]{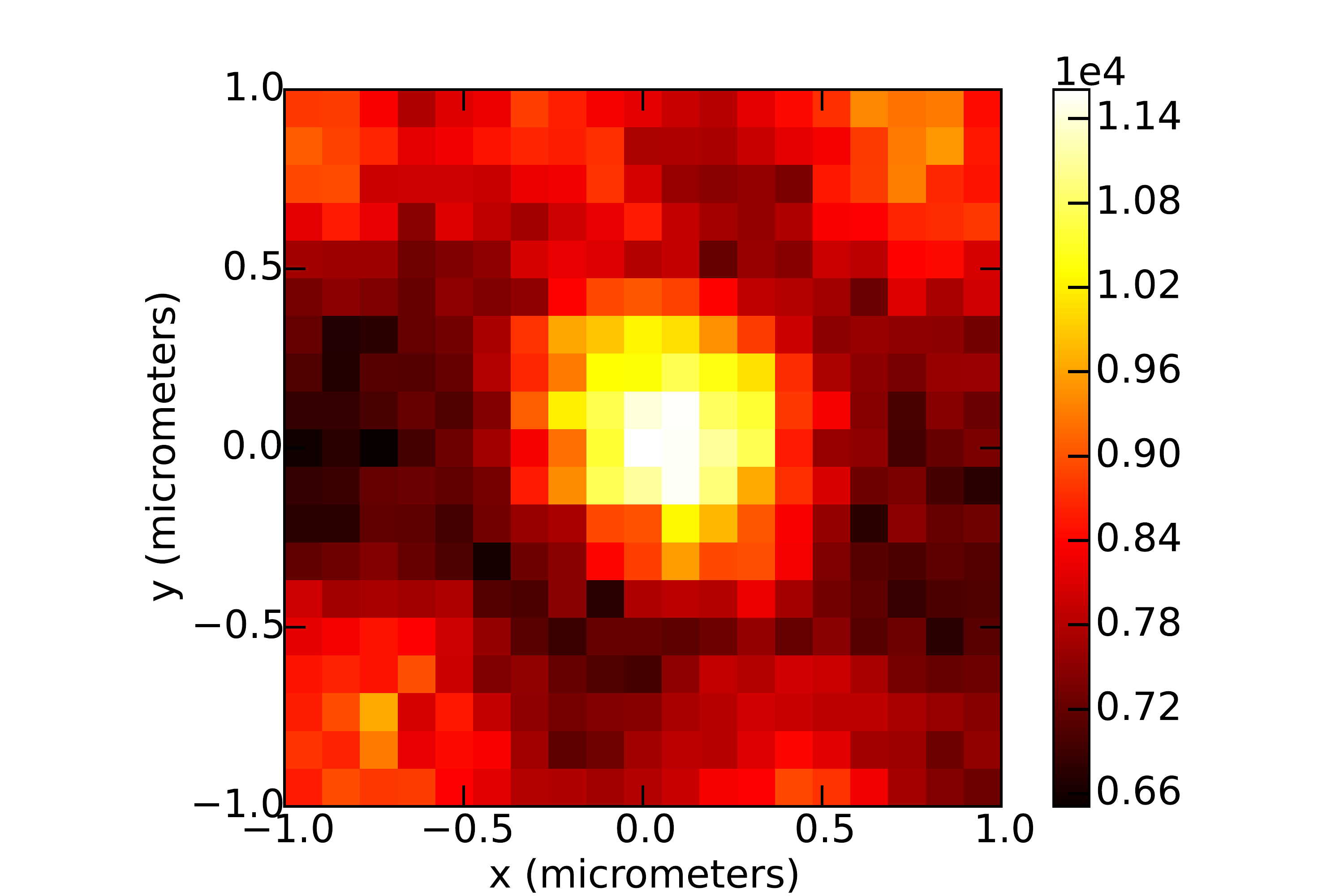}
\caption{\label{fig:plScan} Representative two-photon-excited
photoluminescence image of a single silicon-vacancy defect obtained by scanning the sample 
through the 1040\,nm pulsed excitation laser beam focus and imaging the emitted fluorescence
through a 13\,nm-wide bandpass filter centered at 740\,nm onto the sCMOS camera. Each pixel 
in the image represents the total counts collected by the camera within a digitally-defined 
``pinhole'' approximately ten times larger than the imaging point-spread function. }
\end{figure}

\begin{figure}
\includegraphics[width=8.6cm]{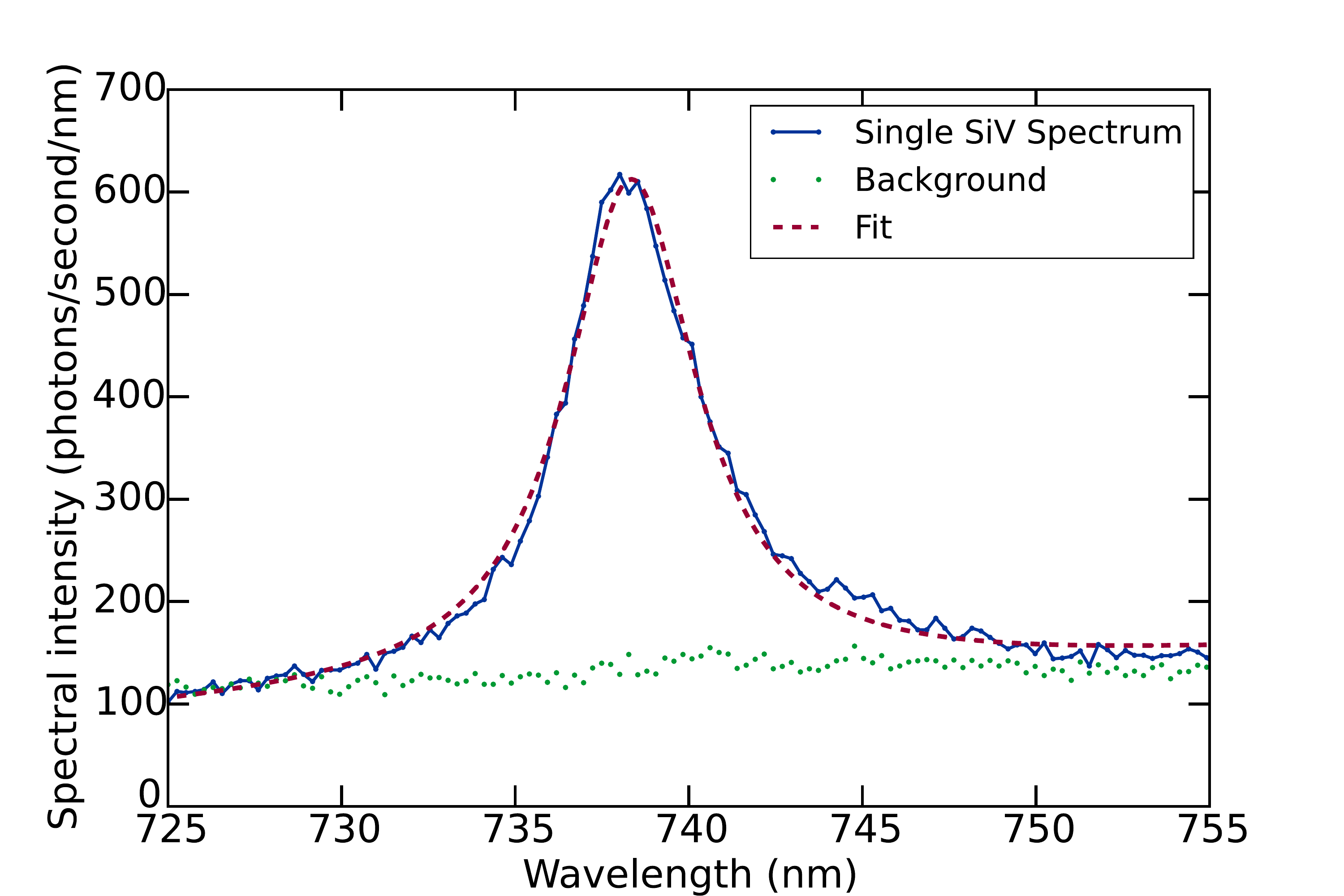} 
\caption{\label{fig:spectrum} Measured spectrum of a single two-photon-excited SiV color center (solid line
and points).
The color center is illuminated by 132\,mW of laser power at 1040nm, with a repetition
rate of 100\,MHz and a pulse duration $2\tau= 144$\,fs (full width at 1/$e$ of maximum).
The spectrometer acquisition time was 
30\,s. 
 A spectrum at a location 2\,$\mu$m distant (dotted line),
  not containing
 any color center, is shown below to indicate  the spectral background.
 The red dashed curve is a fit
to a lorentzian with a second-order-polynomial background. The measured 
width of the SiV ZPL is $4.6\,$nm.
}
\end{figure}

\section{Experimental Apparatus}
The experimental apparatus (illustrated in Fig.~\ref{fig:apparatusSeparate}) 
consisted of a two-photon epifluorescence microscope to excite and detect
 silicon-vacancy color centers in a diamond sample.
Excitation light was derived from a 
mode-locked  1040nm ytterbium  fiber  laser  (Menlo Orange)  producing
$~144\,$fs (full-width  at $1/e$  of maximum)  pulses at  a repetition
rate of 100\,MHz  with an average power of around  300\,mW.
The excitation light was focused via a
$40\times$, NA=0.75 air-spaced objective into a diamond sample mounted
on  a three-axis  closed-loop  translation stage (Newport VP-25XL and 
XPS controller).   Two-photon-excited
fluorescence  from the  sample  was collected  and  collimated by  the
objective,  reflected  from  a  dichroic beam  splitter,  filtered  by
long-pass and  band-pass filters, and  refocused onto an  sCMOS camera
(Hamamatsu Orca Flash4.0) or alternatively free-space coupled into
 a spectrograph (Princeton Instruments Acton SP2500) for spectral analysis of the emitted light. 

We employed samples consisting of bulk diamond implanted with  silicon ions
and annealed to produce silicon vacancies and minimize undesired defects.
High-purity CVD-grown single-crystal diamond chips ($<5\,$ppb nitrogen, Element Six), 
were Si-ion implanted (Materials Diagnostics; Albany, NY),
with the implantation energies chosen to achieve the desired
depth profile, as calculated by the stopping range of ions in matter (SRIM) software \cite{2010SRIM}. 
The implantation surface was a $( 100 )$  plane of the diamond crystal.
Low-density 
samples for observation of single color centers
were fabricated at an implantation density of
$5\times 10^{9}\,\text{ions/cm}^2$ 
and ion energy of 3\,MeV, while high-density samples for improved
signal-to-noise measurement ratio were fabricated
at implantation densities from
$10^{12}\,\text{ions/cm}^2$ to 
$10^{15}\,\text{ions/cm}^2$ with a selection of energies between
500\,keV and 3\,MeV designed to produce approximately uniform SiV
density over the first micrometer of subsurface crystal depth.
After implantation, the samples were annealed in vacuum by ramping over 2\,hrs
to 400\,$^\circ$C and holding for 4\,hrs, then ramping over 2\,hrs to 800\,$^\circ$C
and holding for 4\,hrs, before ramping back to room temperature.

Isolated  single-SiV color  centers  were located  in the  low-density
samples by raster  scanning the sample in the plane  transverse to the
optical axis.   Comparison between one-photon  fluorescence microscopy
of  the   SiVs  with   a  large   digital  confocal   ``pinhole''  and
surface-reflection  depth  scans using  the  same  705\,nm laser  beam
indicated that  the color  centers were within  $\sim 1\,\mu$m  of the
diamond surface, as expected from simulations.  A polarizer was placed
in the imaging path following the dichroic mirror to allow analysis of
the polarization direction of the emitted light, and a half-wave plate
was mounted in a  computer-controlled rotation stage immediately above
the  objective.  As  we observed  no dependence  of the  emitted light
intensity on  the excitation polarization with  the polarizer omitted,
rotating  the wave  plate was  considered equivalent  to rotating  the
transmission axis of the polarizer,  but with the advantage of holding
the  detected polarization  axis  constant  throughout the  downstream
imaging system and spectrometer.


The power of the two-photon excitation beam was measured using a slide-format thermal power meter (Thorlabs S175C)
placed in the position normally occupied by the sample. The excitation point spread functions
for the one- and two-photon beams were measured by scanning a single
 color center through the laser focus
and recording a photoluminescence (PL) map, as shown in Fig.~\ref{fig:plScan}. 
Detection point-spread functions were determined by centering
a color center on the laser focus and recording a long exposure ($\sim 10\,$s) on the camera
using a long-pass filter and a narrow band-pass filter to reject any residual excitation light 
reflected by the dichroic beam splitter.

The pulse width  of the  two-photon excitation  laser was  measured by
means  of  a  home-built   autocorrelator  consisting  of  a  scanning
Michelson interferometer with  a GaAsP photodiode at  its output port;
the  GaAsP bandgap  of 1.98\,eV  being larger  than the  single-photon
energy at  1040\,nm, the  photodiode output  was proportional  
 to the square of the incident power. 
The quantity of glass
traversed by the laser beam before impinging on the sample
was sufficiently small to lead us to expect negligible dispersion and pulse broadening at the sample;
this expectation was confirmed by the experimental 
introduction of a glass component of  comparable thickness  prior to the autocorrelator,
which was not found to produce detectable broadening of the pulse.
The  repetition rate was separately measured
on an oscilloscope using the laser's synchronous radiofrequency output.

We further investigated the spectral dependence of the two-photon cross section
using the signal output of a tunable optical parametric oscillator (Coherent Chameleon Compact OPO).
The  OPO signal 
wavelength was tuned under computer control over the range 1010\,nm to
1550\,nm, and  fluorescence was detected  at
738\,nm.    Making    separate   use    of   the    remotely   tunable
titanium-sapphire pump laser allowed us to extend the excitation range
down  to 920nm,  limited  by our  dichroic  beam-splitter.  For  these
measurements,   we  employed   an  NA=0.95,   100$\,\times$  objective
corrected for use in the IR.   The pulse width of the excitation laser
beam  was   determined  as   a  function   of  wavelength   using  the
autocorrelator,  substituting  a  silicon  photodiode  for  the  GaAsP
detector  for wavelengths  beyond  1300nm.  The  excitation power  was
determined by  measuring it in  transmission using a  large-area thermal
power
meter  placed  directly  beneath  the diamond  sample  and  underlying
microscope slide.  The  analog output of the power  meter was recorded
via  a  data-acquisition  system  for each  measurement,  and  a  dark
power-meter reading was also recorded  at regular intervals during the
measurement sequence by shuttering the excitation beam and waiting for
the  power  meter  reading  to  stabilize.   A  half-wave-plate  in  a
computer-controlled  rotator followed  by a  polarizing beam  splitter
allowed us to vary the excitation power for each wavelength.  To focus
the imaging  system, we introduced a  weak laser beam at  736\,nm into
the excitation path  via a pellicle beam splitter;  the imaging system
was periodically focused on the diamond surface by maximizing the peak
intensity of the reflected 736\,nm beam  on the camera with the pulsed
excitation light  shuttered.  At  all other  times, the  736\,nm laser
beam   was   itself   shuttered.   For   our   excitation-spectroscopy
measurements,  a 50:50  beam-splitter  was employed  in  place of  the
flipper mirror, 
allowing simultaneous measurement of 
the  emission   spectrum   and   the
two-dimensional excitation beam profile.

\begin{figure}
\includegraphics[width=8.6cm]{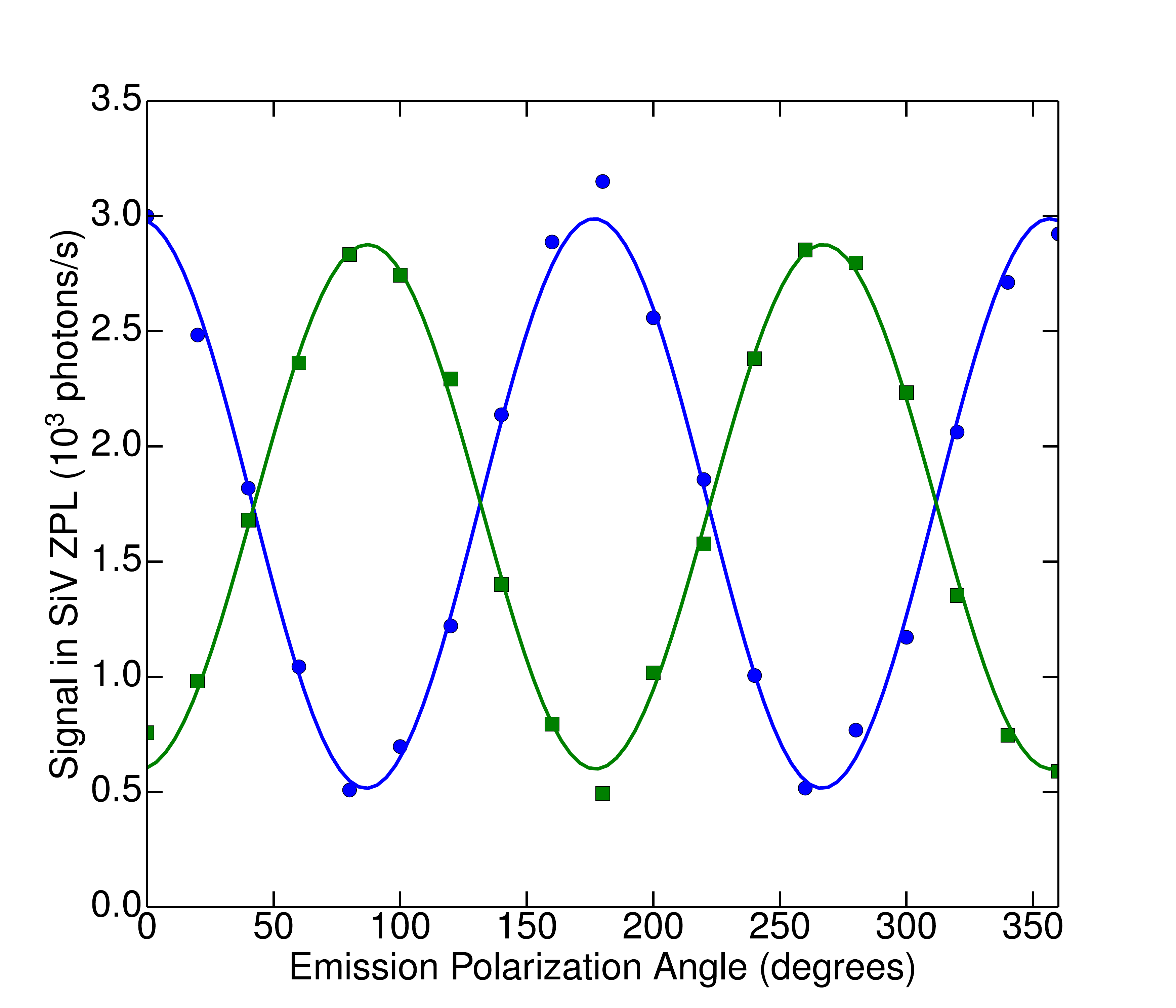}
\caption{\label{fig:polDep}
Polarization dependence of emitted SiV two-photon-excited fluorescence. The data (solid squares and 
circles) are obtained
from two representative color centers, inferred to be oriented along crystal axes with orthogonal 
projections on the image plane. Solid lines are sinusoidal fits to the respective data. The measured
count rate is the area under the ZPL curve, in units of photons per second incident on the spectrometer
camera.
}
\end{figure}

\begin{figure}
\includegraphics[width=8.6cm]{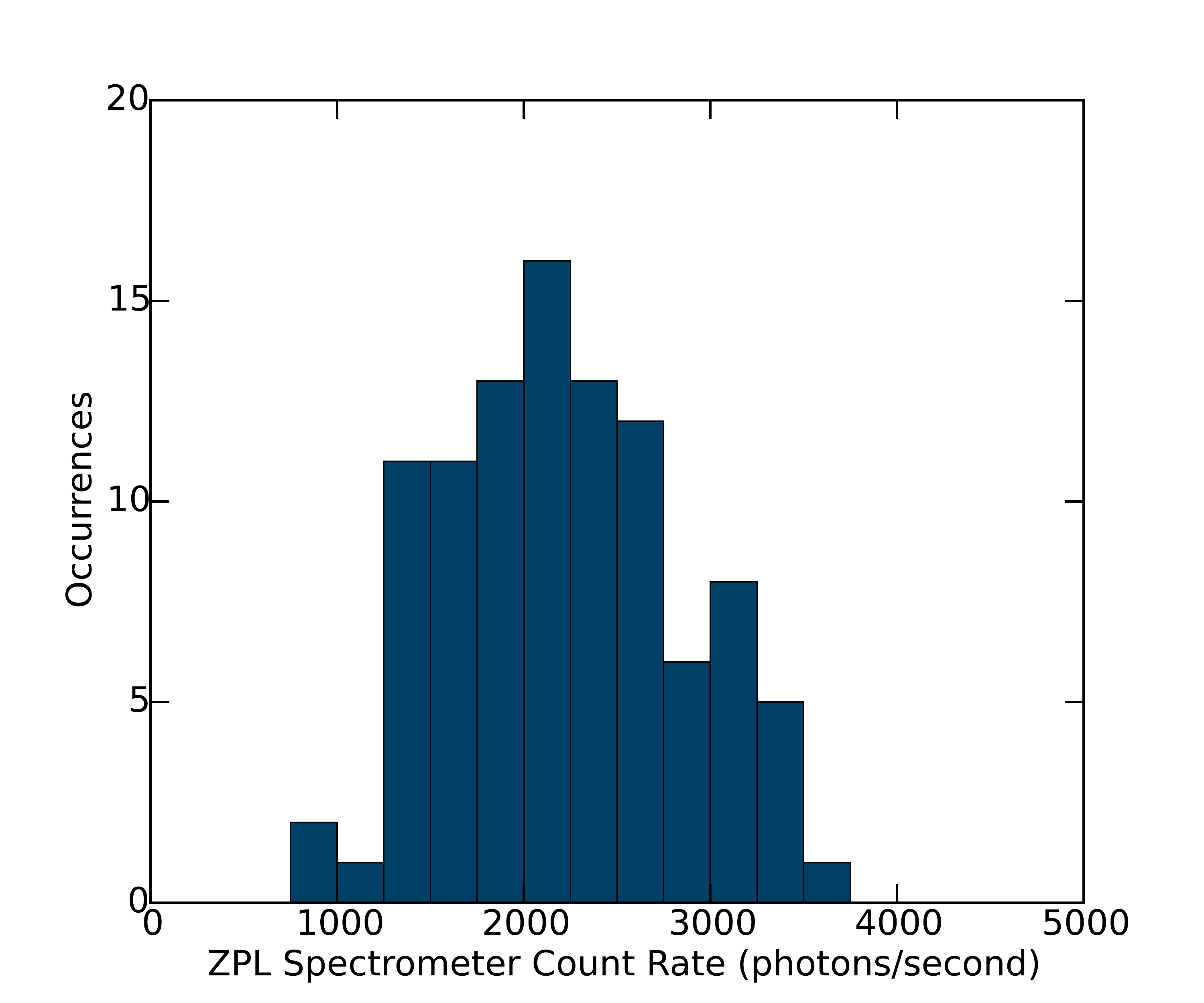}
\caption{\label{fig:brightnessHist} Brightness
distribution of individual SiV color centers obtained from sinusoidally fitting
polarization-dependent spectrometer ZPL count rates. We have
rejected from this distribution  defects with nearly twice the median brightness
or with much lower polarization contrast, presumed to correspond
to double SiVs with (respectively) parallel or orthogonal projected orientations.
}
\end{figure}

\begin{figure}
\includegraphics[width=8.6cm]{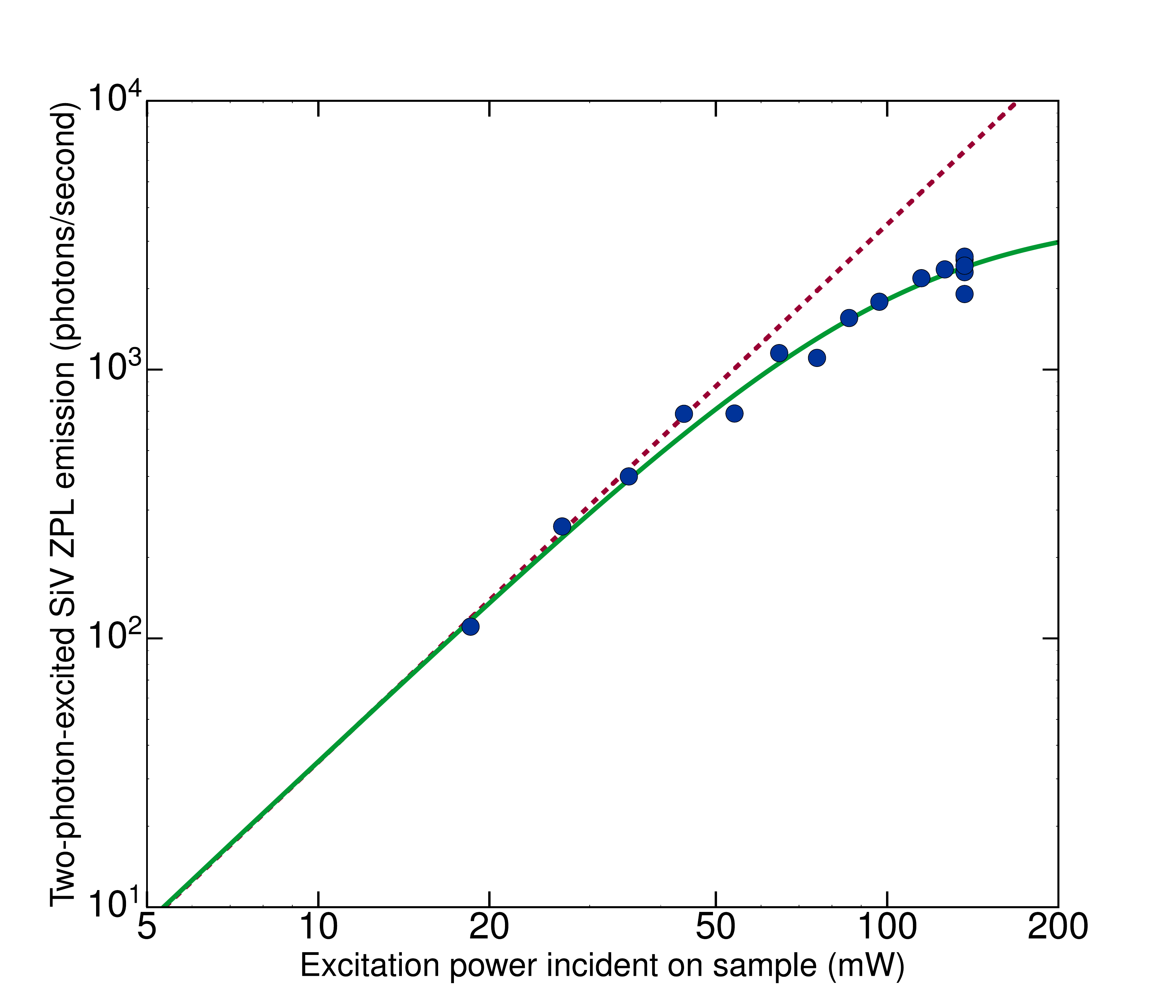}
\caption{\label{fig:powerdep} Power dependence of
the observed two-photon-excited ZPL fluorescence signal
from a single isolated silicon vacancy. 
Solid circles are measured data. The dashed line is a fit to pure quadratic dependence,
while the solid line is the result of a curve fit including saturation.
}
\end{figure}

\begin{table*}
\begin{tabular}{l|l 
|r }
Measured & Value  
&  Partial 
\\
Quantity &   
&  Uncertainty ($10^{-2}$GM) 
\\
\hline
\phantom{\rule{0mm}{4mm}}
$\eta_\text{det.}   $	 
&	\phantom{XX}$4.04(35)\times 10^{-3}$\,$\mathrm{}$	
&		\phantom{XX}6.4	 \\
$\Gamma_\text{det.,2p}$	 
&	\phantom{XX}$2.15(26)\times 10^{3}$\,$\mathrm{s^{-1}}$	
&		\phantom{XX}8.9	 \\
$\tau                     $	 
&	\phantom{XX}$72.0(3.5)$\,$\mathrm{fs}$	
&		\phantom{XX}3.6	 \\
$A_\text{ex.}    $	 
&	\phantom{XX}$8.79(77)\times 10^{-13}$\,$\mathrm{{\rm m^2}}$	
&		\phantom{XX}12.9	 \\
$\langle P \rangle              $	 
&	\phantom{XX}$101.5(3.8) $\,$\mathrm{mW}$	
&		\phantom{XX}5.5	 \\
${S_P}                    $	 
&	\phantom{XX}$0.355(20)$\,$\mathrm{}$	
&		\phantom{XX}4.2	 \\
\hline
Total:                   	 & 
& 		\phantom{XX}19.0	 
\end{tabular}
\caption{\label{table:absolute}Principal contributors to the 
error budget for the absolute measurement of SiV $\sigma_\mathrm{2p}$ with
excitation at 1040\,nm and detection on the ZPL at 738\,nm. The reported error
in each quantity is an estimated single  standard deviation.
The total uncertainty is the sum in quadrature of the partial uncertainties.
Symbols are as defined in the text.
}
\end{table*}

\section{ Results} 
We determined the silicon-vacancy two-photon cross section 
$\sigma_\text{2p}$ via
%
Eq.~\ref{eq:sigma2pMeasurablesSingles},  with  all  terms  apart  from
$\sigma_\text{2p}$ determined from experimentally measured quantities.
In  order   to  obtain  reliable  statistics   on  single-color-center
emitters,  we  performed automated  coarse,  large-area  scans of  the
transverse  ($x$-$y$) position.   Candidate  SiVs  were identified  by
means  of  the  narrow ZPL   at  $\sim738$\,nm (with a median width of $5.6\,$nm FWHM
for our samples),  as  shown  in  Fig.~\ref{fig:spectrum}. When  the spectral  power in a  window surrounding
the ZPL exceeded the background in this window
by  an empirically-determined threshold of 10\%, the position of
the candidate  color center in  the focus  of the excitation  beam was
then optimized  by  sequential  linear  scans  in the  $x$,  $y$,  and  $z$
directions.  When this  optimization  routine terminated  successfully
(i.e., with an above-threshold brightness and at a position not on the
boundary  of  the linear  scans),  the  polarization contrast  of  the
emitted light was then measured by  recording spectra as a function of
the  wave-plate  angle. Polarization  curves  for  162 distinct  color
centers were extracted from the  raw spectrometer data by fitting each
SiV  ZPL  spectrum  to  a lorentzian  with  a  first-order  polynomial
background to obtain the area  under the ZPL spectrum.  Representative
polarization  curves are  shown in  Fig.~\ref{fig:polDep}.  Individual
SiVs may  be aligned along  any of  four allowed axes  (the unoriented
body diagonals of the cubic diamond  unit cell); the projection of the
SiV axis on the plane transverse  to the imaging axis can therefore be
aligned  along two  orthogonal directions,  which for  our sample  are
parallel  to the  edges of  the rectangular  diamond chip  and to  the
principal  axes of  the  detection polarizer.  Consequently we  expect
individual  SiVs to  fall into  two classes,  with orthogonal  emitted
polarizations. Indeed, we observe that  approximately half of the SiVs
identified in a large scan have a maximum detection rate when the slow
axis of the  half wave plate is  set to an angle  close to $0^\circ$,
and approximately half have  a maximum at an angle  around $45^\circ$, corresponding
to  emission  polarizations  at  angles of  $0^\circ$  and  $90^\circ$
respectively. 
The polarization dependence and spectral width are consistent with those
reported for the SiV defect
in other studies \cite{neu2011fluorescence,rogers2014electronic,
wang2006single,vlasov2009nanodiamond}.

The total ZPL photon count rate for  each SiV was then taken to be the
peak-to-peak  amplitude of  a cosine-squared  fit to  the polarization
curve. This count rate displayed  a somewhat wide distribution, as shown
in  Fig.~\ref{fig:brightnessHist}. We  have confirmed  via Monte-Carlo
simulations  that  a  distribution  of  approximately  this  width  is
expected  due  to   fluctuations  in  the  number   of  color  centers
contributing  to the  signal.   Indeed, while the  number  of color  centers
contributing  to  a  single  measurement  is close  to  one,    the
polarization curves in some cases receive  non-negligible contributions from nearby
color centers. Color  centers with parallel dipole  emission axes tend
to enhance the  amplitude of the polarization curve,  while those with
orthogonal  dipole emission  axes tend  to reduce  the contrast.   The
extreme cases of  two color centers very close  together with parallel
or perpendicular emission axes can  be excluded by rejecting candidate
color  centers  whose  polarization   curves  have  exceptionally  low
contrast or  exceptionally large amplitudes  (specifically, amplitudes
close  to twice  the typical  single-SiV amplitude).   After excluding
color  centers with   low  contrast (40 of the 162 candidate centers)
or  anomalously high  amplitudes (20 of the candidate centers), the
median  peak-to-peak amplitude  was  taken  as the  value  of the  ZPL
detection count rate. The uncertainty in this value was dominated by the
uncertainty of the spectrometer intensity calibration, so that in practice
excluding these centers had a negligible effect on our result.

The saturation factor $S_P$ was determined by varying the excitation power
incident on a single SiV color center centered on the laser focus in three dimensions 
 and recording the detected ZPL fluorescence. 
Results of one such measurement
are shown in Fig.~\ref{fig:powerdep}. 
These data were empirically well fit by the 
 functional form
\begin{equation}
\label{eq:satFuncForm}
\Gamma(P) \propto S_P P^2 = \frac{ P^2}{1+(P/P_\text{sat.})^2}
,
\end{equation}
where $\Gamma$ is the detection  rate at power $P$ and $P_\text{sat.}$
is a time-averaged saturation power.  For the measured beam shape, the
fitted  saturation  power corresponds  to  a  saturation intensity  of
$8.3\times 10^{9}\,\text{mW/cm$^2$}$, with a  saturation count rate of
$3.8\times 10^3$\,photons/s incident on the  spectrometer camera.  We point out that
this count rate  is limited in our  system by the fact  that a silicon
vacancy  can  only  emit  photons every  laser  repetition  period  of
$\sim10\,$ns, rather than every excited-state lifetime of $\sim1\,$ns.
Taking  this difference  into account,  our saturation  count rate  is
comparable     to     the     one-photon    SiV saturation     rate     of
$56\times      10^3\,$counts/s     measured      by     Rogers      et
al.~\cite{rogers2014multiple}  for  an   air-spaced  objective.   From
analysis of our saturation  measurements, we determined the saturation
factor at  the power  used for  the cross  section measurements  to be
$S_P=0.36$.  For a  diamond sample containing many SiVs per excitation volume,
the saturation behavior given  by Eq.~\ref{eq:satFuncForm} is modified
because color  centers in the center  of the beam experience  a higher
laser   intensity  and   therefore  saturate   before  those   on  the
periphery. Assuming  a gaussian laser intensity  profile, the expected
saturation behavior for a dense  sample can be calculated analytically
by averaging $\Gamma$ over spatial locations, the result being
\begin{equation}
\label{eq:satFuncDense}
\Gamma(P) \propto P_\text{sat.}^2 \log\left \{ 1+ 
P^2/P_\text{sat.}^2
\right \}.
\end{equation}
The saturation behavior  for a dense sample was also  measured and fit
using  Eq.~\ref{eq:satFuncDense},  and a   consistent  saturation
intensity   was   obtained.     Non-negligible   saturation   of   the
two-photon-excited fluorescence also changes the spatial profile of the
PL map  obtained by scanning the  sample in the $x$-$y$  plane.  At low
intensity,  the photoluminescence (PL)  map  is  expected to  have  a width  approximately
$\sqrt{2}$ smaller than the  actual excitation intensity distribution,
assuming a near-gaussian beam profile,
as a  result of the  quadratic dependence of scattering  on intensity.
For  finite  intensity,  however,  this  factor  is  reduced;  at  the
saturation factor $S_P=0.36$  noted above, the width of the  PL map is
found numerically  to be a  factor of approximately 1.29  smaller than
the  actual beam  width, but  still  reasonably well approximated  by a  gaussian
profile. Analysis of the transverse PSF including this saturation behavior
 indicated a transverse excitation
beam  width approximately  20\%  larger  than the  diffraction-limited
value.

Calibration  of the  detection efficiency  of our  imaging system  was
performed using  light from a  diode laser  at 736\,nm, with  which we
measured the  reflection or transmission coefficients  of each element
in the  imaging beam path.  Intensity calibration of  the spectrometer
camera  at the  operating  temperature of  $-75^\circ$C was  performed
using the  light from the  same laser beam with  separately calibrated
neutral-density filters,  and agreed with the  manufacturer's value to
within approximately 12\%. The collection efficiency of the objective (including
Fresnel reflection from the diamond-air interface) was calculated using
its numerical aperture and the known refractive index of diamond (see Appendix \ref{appendix:collEff}).

Analysis of our measured data yielded
 a value of 
$\sigma_\text{2p,1040nm}=0.74(19)$\,GM
for the two-photon fluorescence cross section of a single SiV at 1040nm,
where the Goeppert-Mayer (GM) is the unit of two-photon cross section,
equal to $10^{-50}\,\text{cm}^4\,\text{s}/\text{photon}$.
The  contributions  of  the  various sources  of  uncertainty  to  our
cross-section measurement are tabulated in Table~\ref{table:absolute}.
The uncertainty of each quantity has been estimated as
the standard  deviation of  repeated measurements, where possible. 
The  uncertainty in
the  intensity calibration  of  the spectrometer  was  limited by  the
uncertainty in the  power of the calibrating laser beam;  this in turn
was estimated by comparing the  measurements of two power meters after
correcting for their wavelength dependence.
The largest contribution
to the uncertainty of the  measurement is from imperfect knowledge of 
the excitation laser beam shape at the sample, whose measurement 
is in turn limited by residual drift between the objective and translation stage. This uncertainty is estimated from the distribution of beam radii
across the ensemble of single color centers, extracted by scanning
each color center through the laser focus and measuring the emitted 
fluorescence. Color centers whose photoluminescence profile was not well fit 
by a single peak were excluded in the determination of this area.

\begin{figure}
\includegraphics[width=8.6cm]{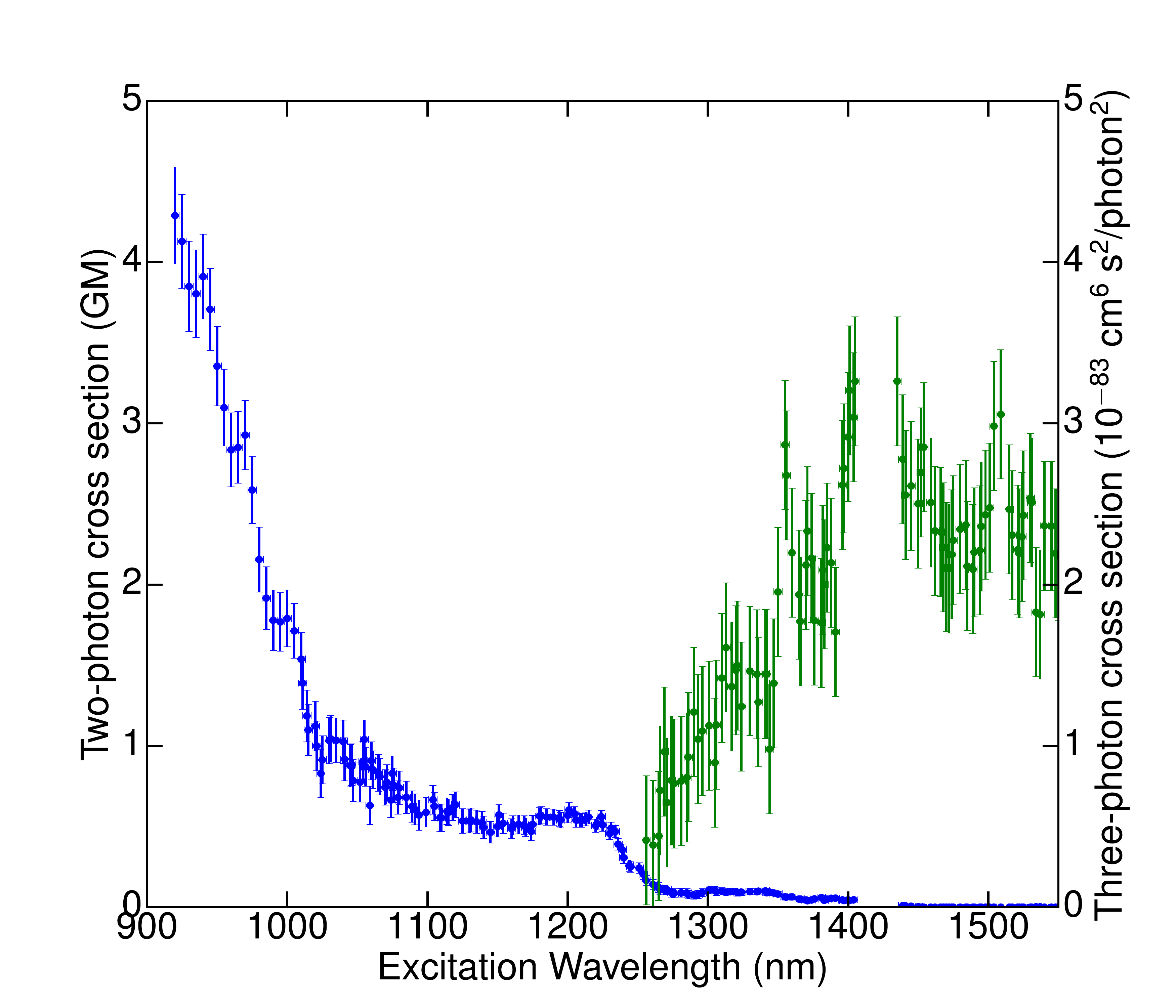}
\caption{\label{fig:excitationSpectra} Two- and three-photon-excited
  fluorescence spectra  (left and  right  axes respectively)  of  a diamond  sample
  densely  populated  with  SiV  color centers.  Vertical  error  bars
  represent statistical uncertainties (standard deviations) of the corresponding
two- and three-photon fluorescence cross sections. A  gap between
1405\,nm and 1435\,nm represents a spectral region in which the excitation power was 
too low for reliable power dependence to be obtained. A multiplicative factor common to
the two- and three-photon spectra has been chosen to  normalize
the two-photon spectrum to the absolutely determined value at 1040\,nm. }
\end{figure}

The results of our excitation-spectroscopy measurements
on a high-density SiV sample  are shown in Fig.~\ref{fig:excitationSpectra}.
For each wavelength,  we determined the photon detection  rate as well
as the two-dimensional  spatial profile of the  sample fluorescence as
functions of  excitation power.  The  photon count rate  was extracted
from the calibrated spectrometer signal by fitting the ZPL signal to a
lorentzian with a first-degree polynomial background.  For wavelengths
below 1300\,nm, the power dependence of  the photon count was fit with
good agreement to a function  of the form of Eq.~\ref{eq:satFuncDense}
to account  for a small  degree of saturation at  shorter wavelengths.
For  wavelengths above  1300\,nm,  this fit  function  no longer  gave
satisfactory  agreement,  and  instead  a  cubic  fit  polynomial  was
required  to obtain  good  agreement with  the  power dependence.   We
interpret  this  as  indicating   that  for  wavelengths  above  $\sim
1300\,$nm,  three-photon  absorption  begins to  make  a  contribution
comparable to that of two-photon  absorption.  At no wavelength in the
range  employed  did  we  observe  significant  linear  dependence  on
excitation  intensity.   The  square  of  the  unit-maximum-normalized
excitation-beam        spatial         profile        $F_\text{ex.}^2$
was   determined    directly   from   the background-subtracted  camera    image   of   the
photoluminescence, allowing  us to calculate the  effective beam areas
$A_2\equiv\int
F_\text{ex.}^2      \text{d}A$,     $A_\text{ex.}\equiv\int      F_\text{ex.}
\text{d}A$,            and           $A_3\equiv\int            F_\text{ex.}^3
\text{d}A$.  The effect  of  saturation  on these  effective  beam areas  was
accounted       for       by               fitting       $A_2$,
                                                             $A_3$,
and                                                     $A_\text{ex.}$
as linear functions of  the  square  of  the  excitation power;  the
saturation-independent  area  in  each  case   was  taken  to  be  the
zero-power intercept of this fit.   From the quadratic and cubic terms
in  our   power-dependence  fit  results,  combined   with  the  other
excitation-wavelength-dependent  factors (excitation  power, effective
beam  areas,  and  pulse  width),  we  extracted  the  two-photon  and
three-photon cross sections shown in Fig.~\ref{fig:excitationSpectra}.
The  product   of  the   two-dimensional  SiV   density  $n_\text{2d}$
and       the       detection       efficiency       $\eta_\text{det}$
at  738\,nm, which  is  common to  the expressions  for  the two-  and
three-photon  spectra, was  chosen to  normalize the  two-photon cross
section to our absolute measurement on single SiVs at 1040nm.

\section{Discussion}

\subsection{Physical Interpretation of the Multiphoton Spectrum}

The excitation spectrum  of Fig.~\ref{fig:excitationSpectra} possesses
a number of  interesting features that hint at  the underlying physics
of the  SiV defect. First,  the spectrum  displays no visible  peak at
twice the  one-photon ZPL  wavelength (1476\,nm).  This  is consistent
with $D_{3d}$ symmetry of the SiV center, which does not permit a pure
two-photon transition  between the $^2E_u$ excited  and $^2E_g$ ground
states. Interaction with parity-odd phonons could still allow enhanced
two-photon absorption near twice the  ZPL wavelength, but our spectrum
does not display any noticeable phonon sideband in the spectral region
below 1476\,nm.
Our excitation spectrum also  shows a considerable increase in
two-photon-excited  fluorescence at  the short-wavelength  end of  the
spectrum. A possible  explanation is excitation to  the continuum, but
\emph{ab initio}  calculations\cite{goss1996twelve} indicate  that the
depth of  the SiV  levels below  the conduction band  is too  great to
explain  the  observed  increase.   Alternative  explanations  of  the
increased  two-photon-excited  emission  at  shorter  wavelengths  are
excitation via another, previously unknown bound level, most likely of
even parity, or  resonant enhancement of the  two-photon absorption by
the  intermediate  $^2E_u$  excited  state.    We also note a  small  bump  in  the
two-photon  excitation spectrum  around  1200\,nm, which  may reflect  direct
two-photon  excitation on  the $^2E_\text{g}$  to the  $^2A_\text{1g}$
transition; a transition between these levels would be  two-photon  allowed even  in  the  absence  of
phonons.  A   thorough elucidation of the  origins of the observed  spectral dependence
and that  of the  observed three-photon  absorption, including  a more
detailed  treatment of  the  role of  symmetry,  will require  further
investigation.



\subsection{Prospects for Deep-Tissue Imaging}

When detecting  a fluorophore  in the presence  of a  large background
signal, such as autofluorescence from biological tissue, the practical detection
limit for a given integration time is  defined by the condition  that the detected signal  from the
label must exceed   the noise from  the background.   Since background
light outside the spectral bandwidth of  the signal of interest can in principle be
rejected  via spectral  filtering,  the relevant shot-noise-limited
background noise is proportional to the square root of the emission line width.
Thus   the shot-noise-limited figure  of
merit $\mathcal{M}$ for   an optical  label  in deep-tissue  molecular imaging  is
approximately given by  the fluorescence cross section divided by the  square root of
the line width,
\begin{equation}
\mathcal{M} = \frac{\sigma_\text{fluor.}}{\sqrt{\gamma}}
.
\end{equation}
Using this figure of merit, we can predict the relative performance of
similar  types  of  labels  for deep-tissue  imaging.   For  instance,
two-photon excitation  of NV  centers in diamond were  measured by  Wee et
al. \cite{wee2007two} to  have a two-photon absorption cross  section of 0.45\,GM,
(corresponding to a two-photon fluorescence cross section of around 0.32\,GM),
and the NV  emission linewidth is approximately  $100\,$nm. 
Our measured  SiV ZPL line width is 5.6\,nm;  Neu et al.\cite{neu2011single}
have observed
room-temperature line widths as small as $0.7\,$nm in SiV-containing nanodiamonds grown on iridium, though such narrow linewidths have not to our knowledge been observed in bulk diamond
or in other nanodiamond samples.
Consequently,
our measurements  suggest that  the figure  of merit  of a  SiV defect
label, with a  two-photon cross section of $0.74\,$GM  at 1040\,nm and
line widths  of 0.7\,nm to  5\,nm, is between  10 and 30  times larger
than  that  of  an  NV  defect,   pointing  to  a  bright  future  for
silicon-vacancy-doped diamond two-photon  labels in applications where
their  photo-  and  chemical  stability are  of  key  importance.  Our
spectral  measurements  indicate  that  the figure  of  merit  can  be
enhanced   further    by   tuning   the   excitation    laser   toward
900\,nm. Furthermore,  the narrowness  of the  SiV ZPL  offers greater
potential for spectral multiplexing in combination with labels such as
the   recently-discovered   narrow-line  germanium   vacancy   defects
\cite{iwasaki2015germanium} or the large number of other known defects
in diamond \cite{jelezko2006single,aharonovich2011diamond}.

\begin{figure}
\includegraphics[width=8.6cm]{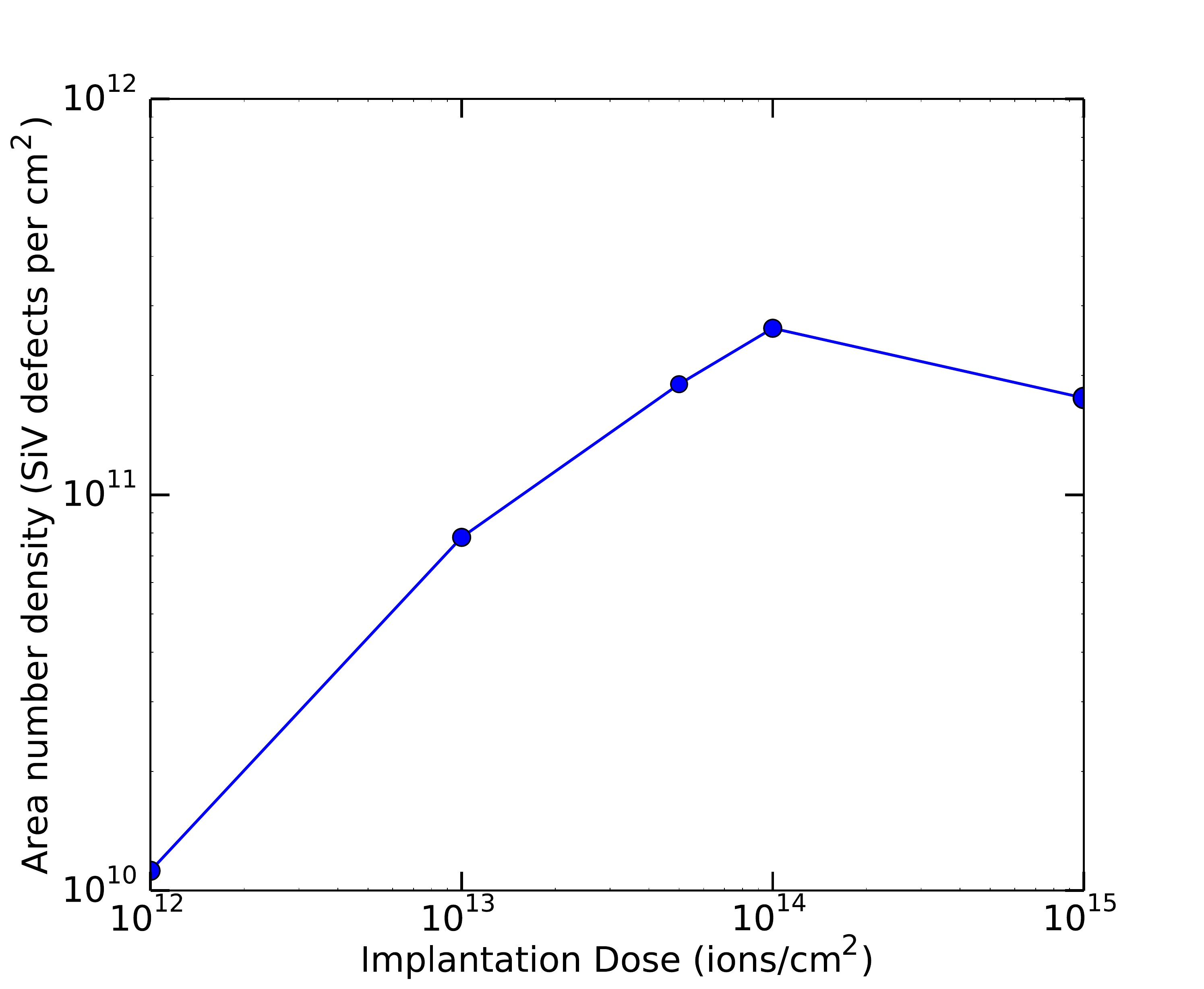}
\caption{\label{fig:implantationDensity}Number density of silicon vacancies  post-annealing
as a function
of the implantation ion dose.}
\end{figure}

Also   important  in   an   evaluation  of   the   prospects  of   SiV
defect-containing  diamond labels  is  an assessment  of  the ease  of
producing  nanoparticle  labels containing  a  high  density of  color
centers.  We have studied the  achievable density of SiV color centers
by systematically varying the implantation  dose and measuring the SiV
one-photon          fluorescence,         as          shown         in
Fig.~\ref{fig:implantationDensity}.
Our      measurements      show      a     maximum      of      around
$n_\text{2d}=3\times  10^{11}\,\text{cm}^{-2}$   in  the   density  of
silicon vacancies produced as a function of the implantation ion dose,
as shown in Fig.~\ref{fig:implantationDensity}.  Assuming based on our
SRIM  simulation  results  that these  vacancies  are  quasi-uniformly
distributed over  a depth  of $1\,\mu$m, the  three-dimensional number
density is then  $n_\text{3d}\approx 3\times 10^{15}\,\text{cm}^{-3}$,
or 17\,ppb.  This  density is substantially ($\sim  10^3$ times) lower
than that achieved for  NV centers \cite{acosta2009diamonds}. However,
work by  d'Hanens-Johansson et al.~\cite{HaenensJohansson2011optical}
has shown  densities of  neutrally-charged SiV centers  with densities
around     $10^{17}\,{\rm      cm^{-3}}$,     while      Vlasov     et
al.~\cite{vlasov2014molecular} have  shown that silicon  vacancies can
exist  stably  in  nanodiamonds  as  small  as  $2\,$nm  in  diameter,
corresponding    to    a    single-nanodiamond   density    of    over
$10^{20}\,{\rm  cm^{-3}}$.  Based  on these  observations, we  believe
that much higher  SiV densities are likely to  be achievable following
additional optimization  of the  growth or implantation  and annealing
steps, allowing realization of bright two-photon SiV nanodiamond labels.

\section{Conclusion}

We  have  measured the  two-photon  cross  section of  the  negatively
charged  silicon-vacancy  color center  in  diamond  at an  excitation
wavelength of  1040\,nm.  Our results yield  an absolute determination
of the two-photon silicon-vacancy cross
section
at 1040\,nm  equal to  approximately 0.74\,GM.   We have  measured the
wavelength dependence  of the  two-photon fluorescence  cross section,
finding a significant  increase of the cross section  for the shortest
wavelengths     measured.       Finally,     we      have     assessed
two-photon-interrogated   SiV-containing   diamonds    as   a   bright
non-bleaching biological label for deep  tissue imaging, and found its
expected  performance  significantly exceeds  that  of  NV centers  in
diamond,  provided  SiV  nanodiamonds  can  be  produced  with  defect
densities  comparable  to  NV  nanodiamonds.  Our  result  shows  that
two-photon-excited  SiV-nanodiamond  labels   offer  an  outstanding
combination  of brightness,  narrow-line emission,  and photostability.
These characteristics promise  improved sensitivity,  greater detection depth,  and longer  interrogation
times in long-term cell tracking and \emph{in-vivo} molecular-imaging experiments,
with significant applications to many topics of fundamental biological
and biomedical interest, including  drug efficacy testing, studies of circulating tumor cells,
and immune cell trafficking.

\appendix

\section{Calculation of Collection Efficiency}
\label{appendix:collEff}
In the  experimental apparatus, the  light emitted by  the fluorophore
must escape from  the diamond substrate before being  collected and transmitted
by the objective.
We compute the fraction of the
total  fluorescence  emitted  by  the color  center  that  enters  the
objective as follows.
\begin{equation}
\eta_\text{coll.}=
\frac{\sum_i\int d\Omega \frac{d\mathcal{F}_i}{d\Omega} T_i}
{\sum_i\int d\Omega \frac{d\mathcal{F}_i}{d\Omega}} 
,\end{equation}
where $d\mathcal{F}_i/d\Omega$  is the fluorescence power  emitted per
solid  angle, $T_i$  is the  Fresnel transmission  probability at  the
diamond-air interface for polarization $i$, and the integration in the
numerator is over the solid angle defined by the numerical aperture of
the  objective.   As  shown below,  the calculated
collection  efficiency  is  the  same,  whether  we  assume  isotropic
emission  or dipole  emission from  a  dipole oriented  along the  $\langle 111 \rangle$
direction of the diamond lattice.

\subsection{Calculation for Isotropic Emission}
If the color center emits isotropic, unpolarized radiation, the collection efficiency 
simplifies to
\begin{equation}
\label{eq:collectionIntegral}
\eta_\text{coll.}=
\frac{1}{4}
\int_0^{\theta_\text{max}} \sin\theta \text{d}\theta 
\left \{T_S(\theta)
+ T_P(\theta) \right \}
,
\end{equation}
where the maximum collection angle is defined by the numerical aperture
of the collection objective, i.e., $n\sin\theta_\text{max}=NA$, $n$ is the index of 
refraction of diamond, and
the transmission coefficients from diamond into air
for S and P polarization are
\begin{equation}
T_S=\frac{
n\cos\theta - \sqrt{1-n^2\sin^2\theta}
}{
n\cos\theta + \sqrt{1-n^2\sin^2\theta}
}
\end{equation}
and
\begin{equation}
T_P=\frac{
n \sqrt{1-n^2\sin^2\theta}-\cos\theta 
}{
n \sqrt{1-n^2\sin^2\theta}+\cos\theta 
}
.\end{equation}
Eq.~\ref{eq:collectionIntegral} can be integrated numerically, using a diamond dielectric 
constant of $n=2.42$ and a numerical aperture NA of 0.75, implying an angle $\theta_\text{max.}\approx 0.315\,$rad., resulting in a collection efficiency 
 $\eta_\text{coll.}\approx 0.0203$.

\subsection{Calculation for Dipole Emission}
The dipole emission amplitude along direction $\hat{n}$ due to a dipole oscillating along
axis $\hat{p}$ is proportional to $\hat{n}\times(\hat{n}\times\hat{p}) = \hat{n}(\hat{n}\cdot \hat{p}) - \hat{p}$.
Here we take $\hat{p}=\frac{1}{\sqrt{3}}\left ( \hat{x}+\hat{y}+\hat{z}\right)$, with the origin
of spherical co-ordinates at the color center and the polar axis normal to the diamond-air 
interface. The two polarization axes are  $\hat{\phi}=-\sin{\phi}\hat{x}+\cos\phi\hat{y}$
and $\hat{\theta}=\cos\theta\cos\phi\hat{x}+\cos\theta
\sin\phi\hat{y}-\sin\theta\hat{z}$, corresponding to S and P
polarization at the interface. Thus the amplitudes for emission with these polarizations
are, respectively, 
\begin{equation}
\mathcal{A}_S=-\alpha\hat{\phi}\cdot\hat{p}
=-\frac{\alpha}{\sqrt{3}}\Big(\cos\phi-\sin\phi\Big)
\end{equation} 
and
\begin{equation}\mathcal{A}_S=-\alpha\hat{\theta}\cdot\hat{p}=
-\frac{\alpha}{\sqrt{3}}\Big(
\cos\theta (\cos\phi+\sin\phi) - \sin\theta
\Big),
\end{equation} 
where $\alpha$ is a proportionality constant, and where we have 
used the fact that $\hat{n}$ is orthogonal to both polarization unit vectors.
The proportionality constant is determined to be $\alpha=\sqrt{3/8\pi}$ by 
normalizing to unit probability.
Thus the total collection efficiency is given by 
\begin{align}
\nonumber
\eta_\text{coll.,dipole}&=
\int_0^{\theta_\text{max}} \sin\theta \text{d}\theta \int_0^{2\pi}\text{d}\phi \cdots\\
&\quad\quad
\Big\{
|\mathcal{A}_S|^2 T_S(\theta)
+
|\mathcal{A}_P|^2 T_P(\theta)
\Big\}
,\end{align}

Simplifying the integral for $\eta_\text{coll., dipole}$, we obtain
\begin{align}\nonumber
\eta_\text{coll.,dipole}&=
\frac{1}{8\pi}
\int_0^{\theta_\text{max}} \sin\theta \text{d}\theta \int_0^{2\pi}\text{d}\phi 
\cdots\\
\nonumber
&\quad\quad\quad\Big\{
\Big(
1+2\cos\theta\sin\phi\cos\phi \cdots \\
\nonumber
& \quad\quad  -2\sin\theta\cos\theta (\cos\phi+\sin\phi)
\Big) T_P(\theta)\cdots\\
& \quad\quad+
\Big(1-2\sin\phi\cos\phi
\Big) T_S(\theta)
\Big\}
.\end{align}
Noting that the transmission coefficients depend only on $\theta$, we can perform the
$\phi$ integral; terms in the integrand proportional to $\sin\phi$,$\cos\phi$, and $\sin\phi\cos\phi$ vanish by symmetry, and we obtain
\begin{align}
\nonumber
\eta_\text{coll.,dipole}&=
\frac{1}{8\pi}
\int_0^{\theta_\text{max}} \sin\theta \text{d}\theta \int_0^{2\pi}\text{d}\phi \cdots\\
\nonumber
&\quad\quad
\left \{T_S(\theta)
+
 T_P(\theta)
\right\}
\\
&=
\frac{1}{4}
\int_0^{\theta_\text{max}} \sin\theta \text{d}\theta
\left \{T_S(\theta)
+
 T_P(\theta)
\right\}
.\end{align}
It is interesting to note that the collection efficiency is identical to the isotropic case, and
that this 
equality obtains only for a dipole oriented at an angle of $\cos^{-1}\left(\frac{1}{\sqrt{3}}
\right)$ to the surface
normal.

For color centers within approximately a half wavelength of the surface, the 
Purcell effect modifies the spontaneous emission rate, and thus the 
detected fluorescence \cite{santori2009vertical}. For our samples,  simulations of the implantation process indicate
a defect depth peaked around $1\,\mu$m, and consequently a negligible Purcell enhancement
of the fluorescence rate.

\begin{figure}
\includegraphics[width=8.6cm]{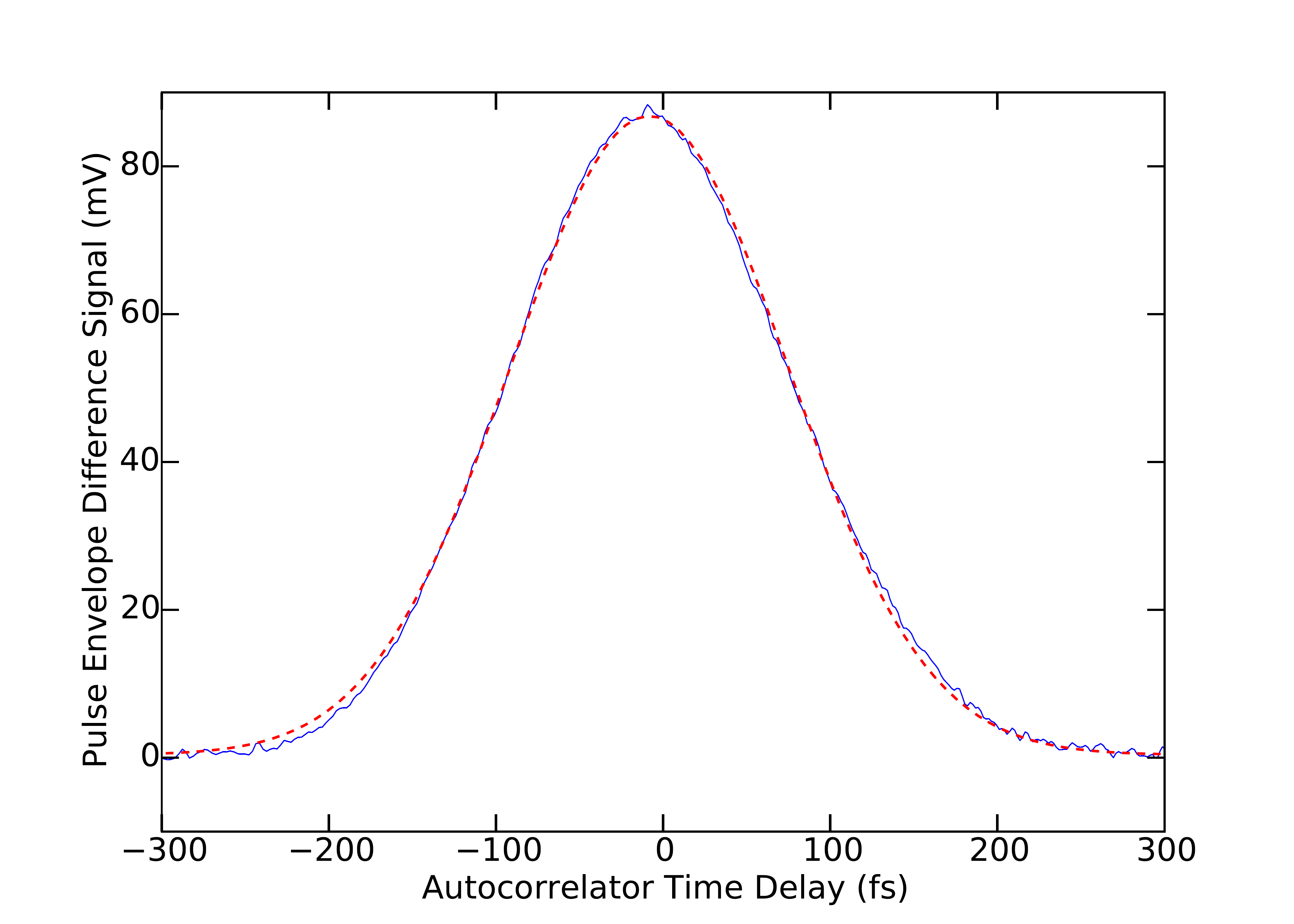}
\caption{\label{fig:autocorrPulse} Autocorrelator  waveform difference
  envelope $\Delta S$.  The  blue  curve  is  obtained  from  the  autocorrelator
  oscilloscope trace by numerically locating  the maxima and minima of
  the  filtered waveform. A   non-uniform interpolation in  time has
been performed to approximately
  linearize the quasi-sinusoidal motion of the interferometer mirror.
The  horizontal axis is calibrated by relating the period of oscillation 
of the waveform to the optical oscillation period. The dashed red curve
is a gaussian fit to the data. The true laser
pulse width $\tau$ is $\sqrt{3/8}$
times the fitted width of this curve, or $72\,$fs.
}
\end{figure}

\section{Pulse Width Measurement}

The pulse width is measured by means of an autocorrelator, as described in Ref.~\cite{diels1985control}. 
In a simple model (e.g., neglecting pulse chirp), the electric field 
for a single pulse incident on the autocorrelator is assumed to 
be of the form 
\begin{equation}
E =  E_0 e^{-i\omega_0 t} e^{-t^2/2\tau^2}
\end{equation}
The electric field at the output port of the Michelson interferometer is then
\begin{equation}
E=  E_0' e^{-i\omega_0 t} e^{-t^2/2\tau^2} +  E_0'e^{i\Delta\phi} e^{-i\omega_0 t} e^{-(t-2\Delta x/c)^2/2\tau^2}
,\end{equation}
where   $\Delta\phi$ 
incorporates the phase difference $2k\Delta x$
due to the mirror displacement $\Delta x$
as well as any static phase difference between the two paths.
The intensity corresponding to this output field is then
of the form
\begin{align}
\nonumber
I&=I_0\Big (
e^{-t^2/\tau^2} + e^{-(t-2\Delta x/c)^2/\tau^2} 
+\cdots\\
\nonumber
&\quad\quad 2\cos(\Delta\phi) e^{-t^2/2\tau^2} e^{-(t-2\Delta x/c)^2/2\tau^2}
\Big) \\
\nonumber
&=
I_0 e^{-t^2/\tau^2}
\Big (
1+e^{4t\Delta x/c\tau^2 - 4\Delta x^2/c^2\tau^2}
+\cdots\\
&\quad\quad \quad \quad 2\cos(\Delta\phi) e^{2t\Delta x/c\tau^2 - 2\Delta x^2/c^2\tau^2}
\Big )
\end{align}
The total signal detected at the photodiode is proportional by design to the square of the 
time-averaged intensity, i.e., to the time-integral of the $I^2$.
Performing this integration yields a photodiode signal $S$, given by
\begin{align}
\nonumber
S&=
S_0 
\Big(
1 + e^{-2 \Delta x^2/c^2\tau^2} +4\cos\Delta\phi e^{-3\Delta x^2/2 c^2\tau^2}
\cdots \\
&\quad\quad \quad\quad  \quad\quad 
+2\cos^2\Delta\phi e^{-2\Delta x^2/ c^2\tau^2}
\Big )
.\end{align}
The upper envelope of the oscillatory curve given by this expression as a function of $\Delta x$
is \begin{equation}
S_+ = S_0(1+3e^{-2 \Delta x^2/c^2\tau^2} + 4e^{-3 \Delta x^2/2c^2\tau^2})
,\end{equation}
while the lower envelope is 
\begin{equation}S_- = S_0(1+3e^{-2 \Delta x^2/c^2\tau^2} - 4e^{-3 \Delta x^2/2c^2\tau^2}.)\end{equation}
Consequently the difference envelope is given by a pure Gaussian,
\begin{equation}
\Delta S \equiv S_+-S_- = 8S_0 e^{-3 \Delta x^2/2c^2\tau^2}.
.\end{equation}
The spatial period of the oscillatory waveform enclosed by these envelopes is $\lambda/2$, corresponding 
to a relative pulse time delay of $\lambda/2c$. 

When this signal is recorded on an oscilloscope as a function of time, the position is assumed to be 
a linear function of time, $\Delta x = \alpha t$. Thus the temporal period $T$ of the oscilloscope trace
is given by $T=\lambda/2\alpha$, or $\alpha=\lambda/2T$.  
The 1/e half-width $t_{1/e}$ of a gaussian fit to $\Delta S$ 
occurs when 
\begin{equation}
 3\alpha^2 t_{1/e}^2/2 c^2\tau^2 = 1
\end{equation}
or
\begin{align}
\tau&=  \sqrt{\frac{3}{8}} \frac{\lambda t_{1/e}}{ cT}
\end{align}
The difference envelope obtained from an
 autocorrelator measurement on our 1040\,nm laser and a gaussian fit to the difference envelope
are shown in Fig.~\ref{fig:autocorrPulse}.

\section{Measurement in Dense Sample}

If the color centers being whose fluorescence is to be collected
are present on a surface with a certain area number density $n_{\rm 2d}(x,y)$,
and the excitation laser beam has a transverse spatial profile given by the function 
$I(x,y)\equiv I_\text{0} F_\text{ex.}(x,y)$, with a spatial
maximum $I_\text{0}$ at the center of the excitation  beam, then the observed signal is given by 
\begin{equation}
\Gamma_\text{det,2p}
=
   \sigma_{2p}  
\int \text{d}x\,\text{d}y\,
\eta_\text{det} G(x,y)
n_\text{2d}(x,y) \langle I^2(x,y)\rangle
,
\end{equation}
where $\eta_\text{det}G(x,y)$ is the spatially-dependent detection efficiency, and 
$G$ is the  detection point spread function (PSF) of the microscope, equal to the convolution
of the confocal pinhole aperture function and the imaging point-spread function at the detection 
wavelength. To avoid ambiguity, we choose $G$ to be a unit-maximum function and to include
the finite on-axis transmission of the pinhole $\eta_\text{pinhole}$ 
as one of several factors contributing to $\eta_\text{det}$. In other words,
\begin{equation}
 \eta_\text{pinhole}G(x,y)\equiv \int \text{d}x'\,\text{d}y'
G_0(x'-x,y'-y)H(x',y'),
\end{equation}
where $G_0$ is the PSF in the absence of any pinhole.
Simplifying under the assumption that the fluorophore density is uniform, 
we obtain
\begin{align}
\nonumber
\Gamma_\text{det,2p}
&=
\eta_\text{det}    \sigma_{2p} \langle I_\text{0}^2\rangle  n_\text{2d} A_2
,\end{align}
where $A_2$ is an effective area defined by the spatial overlap of the
detection point spread function and the square of the excitation point spread function
as
\begin{equation}
A_2\equiv\int \text{d}x\,\text{d}y\,
G(x,y)
F^2_\text{ex.}(x,y)
.\end{equation}
The time average square intensity for a train of gaussian-envelope pulses
is related to the average intensity by Eq.~\ref{eq:timeAverageResult}.
Hence, for such a pulse train, the average detection rate is given by
\begin{equation}
\Gamma_\text{det,2p}=
\eta_\text{det}    \sigma_{2p} \langle I_\text{0} \rangle^2  n_\text{2d} A_2
\frac{T_\text{rep}}{\tau\sqrt{2\pi}}
,\end{equation}

The total excitation power measured by a power meter is related
to the  intensity at the focus by Eq.~\ref{eq:powerIntensityRelation}.
Thus the detected signal is given in terms of the average excitation laser power by
\begin{align}
\nonumber
\Gamma_\text{det,2p}&=
\eta_\text{det}    \sigma_{2p} 
 n_{2d} A_2 
\left(\frac{\langle P \rangle}{A_\text{ex.}}\right)^2 
\frac{ T_\text{rep}}{\tau \sqrt{2 \pi}}
.\end{align}

For three-photon excitation, the calculation is very similar and results in the relation
\[
\Gamma_\text{det,3p} = \eta_\text{det} \sigma_{3p} n_\text{2d} \left(
\frac{ \langle P \rangle}{A_\text{ex.}}\right)^3
\frac{ T_\text{rep}^2 A_3}{\pi \tau^2 \sqrt{3}}
,\]
where \[A_3\equiv \int \text{d}x\, \text{d}y\, G(x,y) F_\text{ex.}^3(x,y).\]



 \end{document}